# In-operando dipole orientation for bipolar injection from air-stable electrodes into organic semiconductors

Anton Kirch[1*], Joan Ràfols-Ribé[1], Kumar Saumya[1], Thushar Salkod Mahabaleshwar[1,3], William Strömberg[1], Ajay Kumar Poonia[1,2,3], Preetam Dacha[1], Yuntao Qiu[1], Sri Harish Kumar Paleti[1,3], Christian Larsen[1], Nicolò Maccaferri[2,3], and Ludvig Edman[1,3]

[1] The Organic Photonics and Electronics Group, Department of Physics, Umeå University, SE-90187 Umeå, Sweden

[2] Ultrafast Nanoscience Group, Department of Physics, Umeå University, SE-90187 Umeå, Sweden

[3] Wallenberg Initiative Materials Science for Sustainability, Department of Physics, Umeå University, SE-90187 Umeå, Sweden

* Correspondence: anton.kirch@umu.se

## Abstract

Efficient charge-carrier injection from air-stable electrodes into organic semiconductors (OSCs) is essential for fabricating solution-processed organic optoelectronic devices under ambient conditions. Today, this is typically achieved by incorporating doped OSC interlayers, introducing self-assembled dipole monolayers, or adding mobile ions to the active material (AM). Here, we demonstrate an alternative approach that eliminates the need for additional injection layers or ionic additives. We achieve this by blending the dipolar compound TMPE-OH into the electroluminescent polymer Super Yellow (SY) and depositing this sole AM between two air-stable electrodes, forming a single-layer, dipole-doped OLED (D-OLED). By tracking its transient voltage-luminance response, performing impedance spectroscopy, and comparing these characteristics with two other single-layer device concepts, i.e. a neat-SY OLED without a dipolar compound and a light-emitting electrochemical cell (LEC) containing mobile ions, we can establish that the auxiliary dipoles in the D-OLED reorient under the applied driving voltage, enabling immediate luminance turn-on and lowering the injection barriers at both electrodes. Finally, we demonstrate that the D-OLED achieves current efficacies comparable to those of SY OLEDs incorporating dedicated injection layers or LECs. Our study establishes dipolar doping as a practical strategy for efficient bipolar charge injection from air-stable electrodes in solution-processed organic semiconductor devices.

## Introduction

Organic semiconductors with tailored properties have enabled the rapid development of organic devices like light-emitting diodes (OLEDs),[1,2] photovoltaics (OPVs),[3,4] transistors,[5,6] photodetectors (OPDs),[7] or sensors.[8,9] These concepts are attractive because they combine high performance with mechanical flexibility,[10] biocompatibity,[11] and offer pathways towards low-cost fabrication from solution under ambient conditions.[12,13] Considering device processability and stability, it is therefore desirable to use air-stable materials for achieving charge-carrier injection into the OSC. This is typically realized by introducing dedicated interface layers, such as electronically doped OSCs[14,15] and dipolar self-assembled monolayers (SAMs),[16,17] or mobile ions into the bulk of the AM.[18] The inclusion of an extra layer, however, exacerbates material recycling and complicates device fabrication under ambient conditions, or even prohibits it when the included material is air-sensitive. Dipolar SAMs further pose the risk that the dipoles reorient in a direction that inhibits charge transfer.[19,20] The mobile-ion approach is limited by its dependence on salt dissolution and ion migration through the bulk to form injection-facilitating electric double layers (EDLs).[21] Such long-range transfer of bulky ions can cause detrimental morphological changes and electrochemical (EC) side reactions.[22]

In this context, recent attempts to polarize the AM by an electric field (E-field) during fabrication are interesting. In 2024, Cui and co-workers reported that the morphology of the AM in OPVs can be tuned by applying a high external E-field in the wet solution state during ambient-air coating, resulting in an improved power conversion efficiency.[23] Zhao et al. reported on a similar wet orientation of ion-transporting monomers in an applied E-field, followed by photochemical curing to freeze the dipole orientation, for the construction of solid-state batteries with vertically aligned ion-transport pathways.[24] Also in 2024, Rodriguez-Lopez and colleagues used molecular dynamics simulations to establish the conditions under which dipolar OSC molecules orient in the direction of an E-field applied during vapor deposition.[25] Finding the optimal molecular orientation state, however, is nontrivial. For instance, light extraction from OLEDs and absorption in OPVs benefit from a horizontal (parallel to the electrode surface) orientation of the transition dipole moment, while efficient charge injection and extraction require vertically oriented permanent dipole moments.[26]

Here, we present an alternative generic approach for efficient charge transfer between two air-stable electrodes and a solid single-layer AM. It builds on the idea by Hofmann et al. to dope polar molecules into a non-polar host to facilitate charge injection, referred to as dipolar doping.[20] They investigated dense films fabricated by thermal evaporation and found a built-in (static after fabrication) impact of the resulting giant surface potential on the hole injection. By contrast, we use a solution-processed polymer host complemented with an auxiliary, electronically insulating dipolar compound. This configuration yields polar reorientation of the dipolar compound by the low applied driving voltage and thus dynamic (the effect is only present under device operation) bipolar injection while keeping the OSCs orientation unaltered. The presented concept is compatible with solution-based fabrication under ambient conditions, does not incorporate mobile ions that can compromise the device performance, does not require high voltages under fabrication, and avoids additional interlayers that increase the stack complexity.

## Results

### Conception

In this work, we aim to understand the functionality of dipolar doping by fabricating a dipole-doped single-layer OLED (D-OLED) and comparing its performance with two other single-layer device concepts, i.e. a neat-SY OLED (N-OLED) and an LEC. The N-OLED comprises neither injection- nor transport-facilitating additives, thus representing the most primitive single-layer device. The LEC contains mobile ions that enable both low injection and transport resistance, owing to EDL formation and EC doping, without the need for additional layers.

Figure 1(a) displays the device structure used for all devices throughout this work. It comprises a glass substrate, a transparent indium tin oxide (ITO) anode (145 nm thick), a single AM film (thickness $d_{AM} \approx 120$ nm), and a reflective Al top cathode (100 nm). The pixel dimensions are 2 x 2 mm². Figure 1(b) presents the work functions of the electrode materials[27] and the highest occupied molecular orbital (HOMO) and lowest unoccupied molecular orbital (LUMO) levels of the employed organic materials, as determined by cyclic voltammetry, cf. Supplementary Information (SI) Section S1. The chemical structures of the AM constituents are shown in Figure 1(c): the electroluminescent copolymer Super Yellow (SY) as OSC, the dipolar hydroxyl-capped trimethylolpropane ethoxylate (TMPE-OH), and the salt $KCF_3SO_3$.

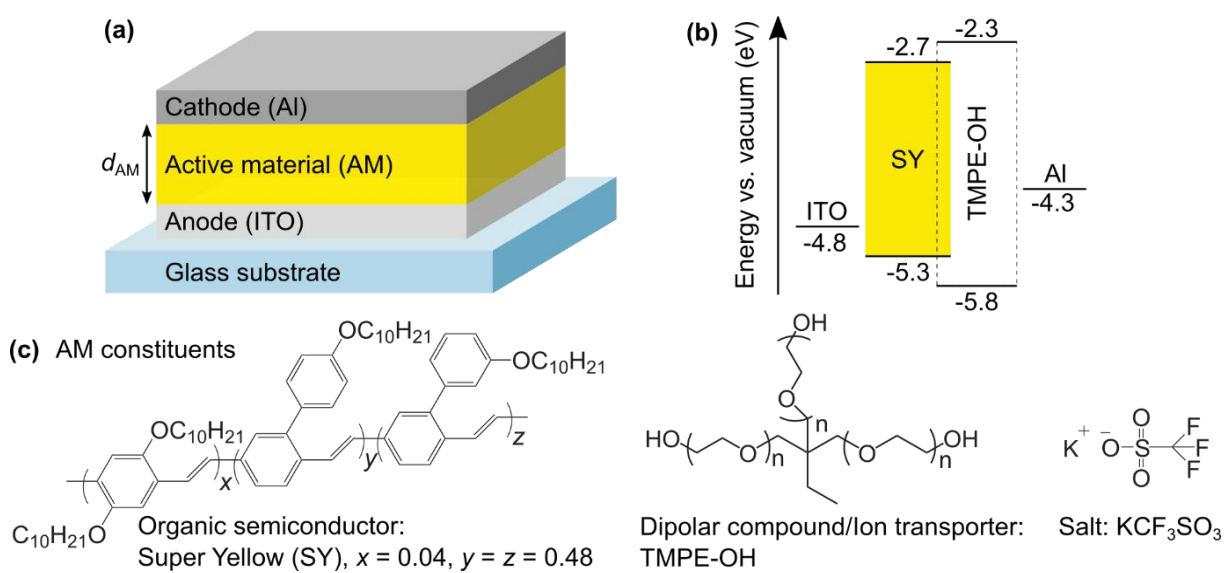

*Figure 1. (a) Sketch of the device structure used in this work. (b) Work functions of the electrode materials[27] and HOMO/LUMO levels of SY and TMPE-OH, cf. SI Section S1. (c) Chemical structures of the AM constituents.*

The three different single-layer electroluminescent device concepts are realized by varying the constituents of the AM. We fabricate

i) a neat-SY OLED (N-OLED), where the AM contains only SY,
ii) a dipole OLED (D-OLED), the subject of this study, comprising SY and TMPE-OH,
iii) three LECs, where the AM consists of SY, TMPE-OH, and salt at three differing salt loads, referred to as low (L), medium (M), and high (H) salt LEC.

Please refer to Table 1 for the specific mass ratios and device abbreviations. We use a TMPE-OH concentration of 40 wt% compared to SY, which is higher than what we usually use for LECs,[28,29] to stay consistent with the optimum TMPE-OH concentration found for the D-OLED.

*Table 1. AM constituents and mass ratios used for the five devices employed in this work.*

| Device | | Abbreviation | Mass ratio | | |
|---|---|---|---|---|---|
| | | | Super Yellow | TMPE-OH | $KCF_3SO_3$ |
| **Neat-SY OLED** | | N-OLED | 1 | --- | --- |
| **Dipole OLED** | | D-OLED | 1 | 0.4 | --- |
| **LEC** | Low salt | L-LEC | 1 | 0.4 | 0.005 |
| | Medium salt | M-LEC | 1 | 0.4 | 0.03 |
| | High salt | H-LEC | 1 | 0.4 | 0.05 |

## Voltage-luminance transients

Figure 2 presents the transients of (a) the driving voltage and (b) the forward luminance for the five pristine devices operated at a constant current of 0.31 mA (corresponding to a current density of 7.7 mA cm$^{-2}$). The N-OLED exhibits a high driving voltage of about 20 V and a faint, constant luminance below 10 cd m$^{-2}$. This is the expected behavior of a simple polymer OLED, with its high driving voltage caused by the large injection barriers, cf. Figure 1(b), and the low electron and hole conductivity of undoped SY.[30] We attribute the poor luminance to an unbalanced charge-carrier distribution, owing to the mismatched injection barriers, i.e. 1.6 eV for electrons vs 0.5 eV for holes.

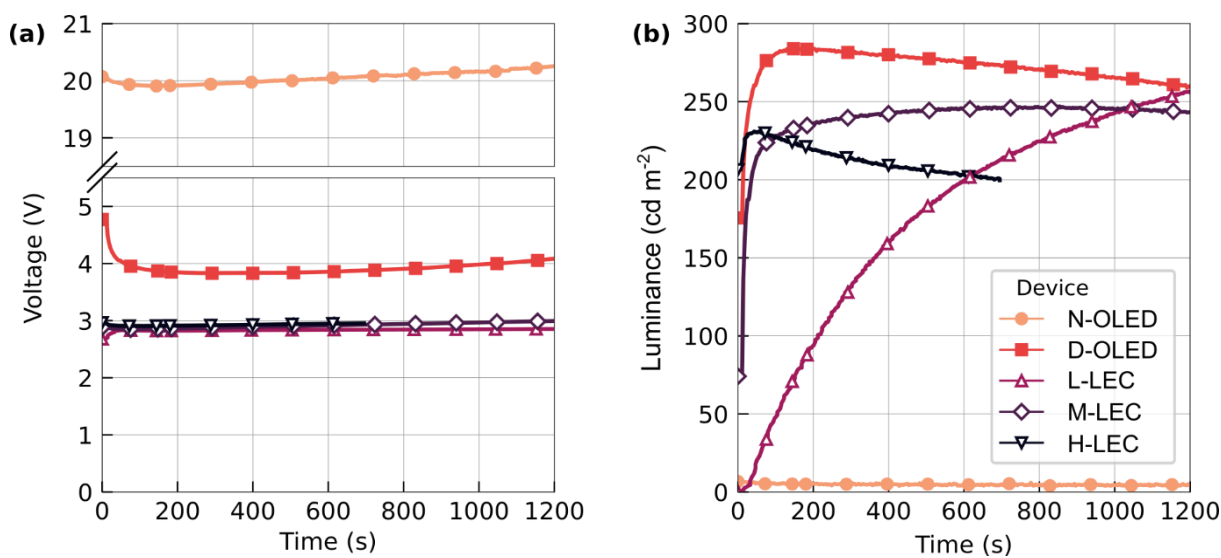

*Figure 2. Temporal evolution of (a) the driving voltage and (b) the forward luminance for the five devices driven at a constant current of 0.31 mA (7.7 mA cm$^{-2}$) using a voltage compliance of 21 V.*

The three LECs (open symbols in Figure 2) reach a driving voltage slightly below 3 V within a few seconds, which is close to the EC energy gap of SY ($E_{SY}$ = 2.6 eV), cf. Figure 1(b). This indicates low injection and transport resistance. In other words, the LECs quickly form EDLs and EC-doped layers, operating close to their theoretical voltage minimum. As expected, their luminance increases during the initial operation. Increasing the salt concentration decreases the turn-on time to peak luminance but also induces a faster onset of the luminance drop. The latter can be assigned to doping-induced exciton quenching and/or electrolyte-induced device degradation. These findings are in agreement with the established LEC understanding.[28,31]

The behavior of the D-OLED is surprising, considering that it features the same injection barriers as the other four devices. Although it is free of mobile ions, it exhibits voltage and luminance transients that indicate changing electrical properties of the AM. The driving voltage drops initially and reaches a minimum of 3.8 V after 340 s, which is a 16 V reduction compared to the N-OLED. Moreover, it turns on quickly, reaching 280 cd m$^{-2}$ after 170 s. This value is comparable to the LECs' luminance and 50 times brighter than the N-OLED.

Figure S2 presents a zoom into the first seconds of operation for pristine pixels of the same devices as in Figure 2, driven at 1 mA (25 mA cm$^{-2}$). Here, we use different experimental equipment that allows us to track the turn-on characteristics at higher time resolution. We can see that while the D-OLED initially requires a high driving voltage of about 14 V, it reduces to moderate voltages below 6 V within 2 s.

Considering the transient behavior of the D-OLED during the initial operation, it is possible to consider ionic impurities in the TMPE-OH as a potential cause. However, the supplier of TMPE-OH reports that the upper limit of potential ionic impurities is below 100 ppm,[32] which corresponds to a ~16 times lower ion concentration than in the L-LEC. Following the salt concentration trend observed in Figure 2(b), such an ion concentration would yield a dramatically slow device turn-on of the D-OLED compared to the LECs, which is not the case. It thus seems unlikely that ionic impurities could be the cause of the D-OLED behavior.

We can deduce that charge-carrier injection and/or transport in the D-OLED works more effectively than in the N-OLED after only a few seconds, albeit it does not reach the electrical performance of the LECs. Further, its charge-carrier distribution seems to be more balanced compared to the N-OLED, allowing electron injection and exciton recombination away from the electrode interfaces, resulting in significant forward luminance and fast turn-on. Both the voltage and luminance transients indicate that the D-OLED's AM properties are not static, like in a classical p-i-n OLED.[33] This is surprising, as no mobile ions are present in the system, and raises the question of how the dipolar compound acts under an applied E-field. To better understand this, we investigate the dynamic processes in all devices via impedance spectroscopy (IS).[34]

## Impedance spectroscopy at short-circuit conditions

Here, we perturb pristine devices with a small AC voltage ($V_{AC}$, RMS = 20 mV), using a DC bias $V_{DC}$ = 0 V at a stabilized temperature, $T$ = 25 °C. The AC frequency $f$ is swept from 100 kHz to 10 mHz and back to 100 kHz, with a full sweep taking about 40 min.

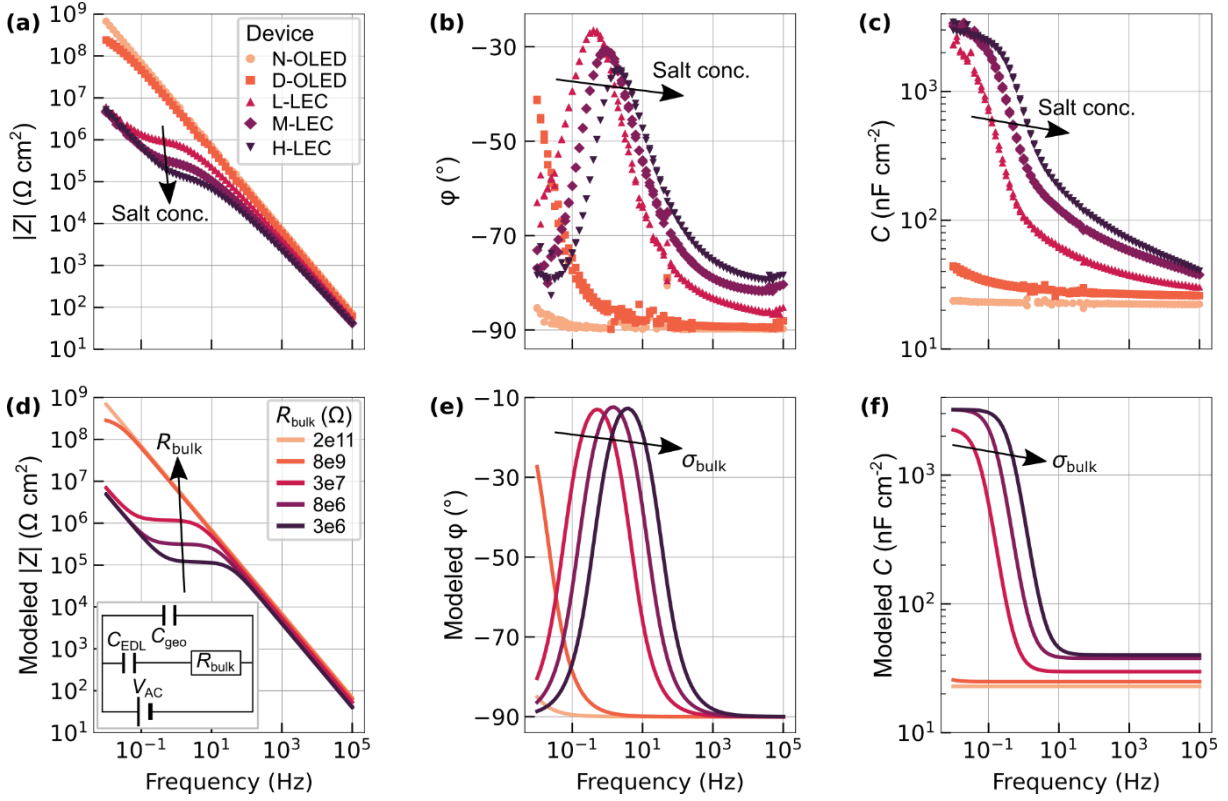

*Figure 3. Impedance characteristics obtained from (a-c) the measurement and (d-f) the equivalent circuit model (ECM) shown in the inset in (d). The data are taken at $V_{DC}$ = 0, with a $V_{AC}$ RMS = 20 mV at T = 25°C. The legend in (a) specifies the tested devices, and the legend in (d) shows the $R_{bulk}$ used as the free parameter in the ECM.*

Figures 3(a,b) present back-and-forth scans of the IS data (absolute impedance $|Z|$ and current-to-voltage phase shift $\varphi$). For all devices, no significant forward-backward hysteresis is visible, indicating that they are measured at steady conditions. Figure 3(c) shows the same data, but represented as the effective capacitance density

$$C = \frac{1}{A\,\omega}\text{Im}\left(\frac{1}{Z}\right) = \frac{-X}{A\,\omega|Z|^2}, \quad (1)$$

with the reactance $X = \text{Im}(Z)$, the pixel area $A$ = 4 mm$^2$, and the angular frequency $\omega = 2\pi f$, cf. SI Section S3.

The N-OLED acts like a capacitor over the investigated frequency range. That is, $|Z|$ scales inversely with $f$, and $\varphi$ is constant at -90°. Using the formula for the parallel-plate capacitance density

$$C = \frac{\epsilon_0 \epsilon_r}{d} \quad (2)$$

and the experimentally determined values $C$ = 23 nF cm$^{-2}$ at 100 kHz (Figure 3c) and $d = d_{AM}$ = 120 ± 5 nm (Table 2), we can extract $\epsilon_r$ = 3.1 ± 0.1 for SY. This value is in agreement with previous reports.[35,36] The capacitance given by the geometric dimension $d_{AM}$ and the relative permittivity $\epsilon_r$ at high frequencies is commonly referred to as geometric capacitance density $C_{geo}$. Here, we use the values found at the upper limit of the investigated IS frequency window, $C_{geo} = C(f$ = 100 kHz).

The D-OLED shows two distinctions from the N-OLED: First, $C_{geo}$ increases slightly (Figure 3c), implying a larger $\epsilon_r$, which can be attributed to the polar TMPE-OH. Second, the phase angle increases at lower frequencies, reaching $\varphi \approx$ -40° at 10 mHz (Figure 3b). This finding is indeed interesting, as it points to a slow relaxation process in the AM. It can be rationalized by TMPE-OH being a polar molecule, featuring three polar OH end groups and an ester backbone that can reorient in an E-field. Such segmental chain reorientation under AC perturbation has been studied in similar molecules like poly(ethylene glycol) (PEG) or poly(ethylene oxide) (PEO).[37,38]

*Table 2. Device parameters for the 5 investigated devices. $d_{AM}$ is measured with a profilometer, $C_{geo}$ and $C_{EDL}$ are the values measured at the indicated frequencies via IS. From $d_{AM}$ and $C_{geo}$, we derive $\epsilon_r$ via Eq. (2). From the ECM, cf. Figure 3(d), we determine $R_{bulk}$ (here displayed as $\sigma_{bulk}$), used as the only fitting parameter. The indicated uncertainties originate from the $d_{AM}$ variations of ± 5 nm.*

|  | Measured | | | Derived | |
|---|---|---|---|---|---|
| **Device** | $d_{AM}$ (nm) | $C_{geo}$ (nF cm$^{-2}$) at 100 kHz | $C_{EDL}$ (nF cm$^{-2}$) at 10 mHz | $\epsilon_r$ at 100 kHz | $\sigma_{bulk}$ (S m$^{-1}$) |
| **N-OLED** | 120 ± 5 | 23 | --- | 3.1 ± 0.1 | < 1.5e-13 |
| **D-OLED** | 128 ± 5 | 25 | --- | 3.6 ± 0.2 | 4e-12 |
| **L-LEC** | 128 ± 5 | 30 | 2300 | 4.3 ± 0.2 | 1.1e-9 |
| **M-LEC** | 124 ± 5 | 37 | 3200 | 5.3 ± 0.2 | 3.9e-9 |
| **H-LEC** | 106 ± 5 | 40 | 3200 | 4.8 ± 0.2 | 8.8e-9 |

Regarding the LECs, introducing mobile ions to the AM clearly changes the IS characteristics and illustrates that their device physics is fundamentally different from the D-OLED. At high frequencies, $C_{geo}$ is increased compared to the N-OLED, as the ions increase the dielectric response of the AM. At $f_c \approx$ 1 Hz, all LECs exhibit a characteristic $\varphi$ peak and a $|Z|$ plateau, cf. Figure 3(a,b). At this point, the ionic conductivity

$$\sigma_{bulk} = \frac{d_{AM}}{A \cdot R_{bulk}} \propto f_c \qquad (3)$$

governs the device impedance.[39] At frequencies $f < f_c$, the three impedance curves collapse back into a single line. Here, the ions have enough time to accumulate at the electrode interfaces, and the EDL capacitance becomes dominant.[40]

To gain a better understanding of the measured data, we use a Debye equivalent circuit model (ECM), cf. inset in Figure 3(d), to reproduce all five device characteristics. This is a common approach to determine the ionic conductivity in polymer electrolytes for the case of blocking electrodes.[41] It comprises merely three elements in its simplest form: The geometric capacitance $C_{geo}$, the EDL capacitance $C_{EDL}$, and the resistance (or conductivity) of the AM bulk, here termed $R_{bulk}$ ($\sigma_{bulk}$), cf. Equation (3). We neglect an additional series resistance, because the impedance of our samples exceeds the resistive contributions of the wiring and electrodes. Modeling the EDL as a capacitor and omitting the use of a Randles circuit is reasonable at $V_{DC}$ = 0, as electronic charge carriers cannot be injected into the OSC (non-Faradaic system), considering their significant injection barriers, cf. Figure 1(b).[34] We choose to model all five devices with the same ECM to achieve a more intuitive basis for comparison. For the N-OLED and D-OLED, the implementation of $C_{EDL}$ is unreasonable, as they do not comprise mobile ions and hence reach no EDL capacitance plateau at low frequencies (cf. Figure 3(c), indicated by "---" in Table 2). This is reflected in the ECM by setting their $C_{EDL}$ to infinity (high values in the implementation). The full set of ECM parameters can be found in the SI Section S4.

As presented in Figure 3(d-f), the IS measurements can be captured qualitatively, and mostly even quantitatively, by this simplistic ECM for all five devices. $C_{geo}$ dominates the impedance at high frequencies and is set equal to the measured $C_{geo}$ (cf. Table 2). When approaching the characteristic frequency $f_c$ for the LECs, the ion or bulk resistance $R_{bulk}$ limits the current.[39] For better comparison, Table 2 displays the respective geometry-independent conductivity values $\sigma_{bulk}$, connected to $R_{bulk}$ via Equation (3). In the ECM, $R_{bulk}$ is the only free parameter and is chosen such that the modeled IS characteristics resemble the measured data. At low frequencies ($f < f_c$), $C_{EDL}$ limits the current for the LECs and is set equal to the measured value, cf. Table 2. For the N-OLED and D-OLED, $C_{EDL}$ is set to infinity (a random high value).

While all LECs show a clear $C_{EDL}$ plateau, no EDL formation is detected for the N-OLED and D-OLED. This is consistent with the absence of mobile ions in these systems. Furthermore, $\sigma_{bulk}$ of the D-OLED, which we can extract from the ECM as the only free parameter, is about 3 orders of magnitude smaller than for the LECs. This supports that the impurity ion concentration is insignificant in the D-OLED. It exhibits, however, a slow relaxation process at low frequencies. This stands in contrast to the N-OLED, which displays a capacitor-like behavior, and indicates that TMPE-OH dipoles slowly reorient in the AM, giving rise to a motion of electric charge.

## Impedance spectroscopy under bias

In the previous section, we probed the devices using impedance spectroscopy at $V_{DC} = 0$ and observed a dipolar relaxation process in the D-OLED. Now, we compare the impedance response of an LEC (the M-LEC as an example) with the D-OLED under increasing bias to learn more about their characteristics under operating conditions. We first let the device stabilize for 20 min at a given $V_{DC}$ and then perform an IS scan with $V_{AC}$ superimposed on $V_{DC}$. The impedance is measured on pristine devices, with the $f$ running from 100 kHz to 100 mHz and back to 100 kHz. Again, we observe no significant forward-backward hysteresis. For $V_{DC} >> 0$ V, electronic charge carriers are injected into the system, and we can no longer treat it as non-Faradaic. Hence, the ECM presented in Figure 3(d) does not hold for this investigation.

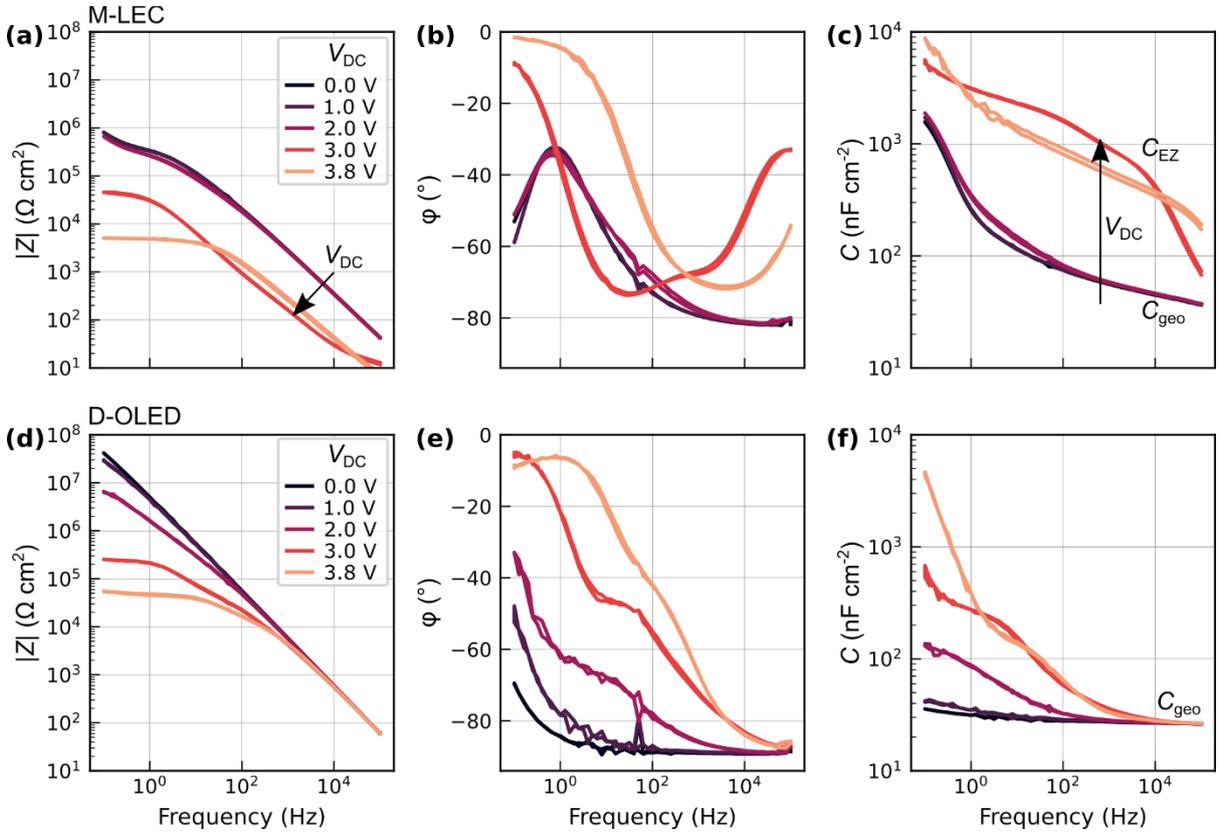

Figure 4. IS data of (a-c) the M-LEC and (d-f) the D-OLED dependent on $V_{DC}$. We use $V_{AC}$ RMS = 20 mV and probe at $T = 25°C$.

Figure 4 compares the IS data of (a-c) the M-LEC with (d-f) the D-OLED. For the LEC, the impedance characteristics ($C_{geo}$, $C_{EDL}$, and $f_c$) remain constant for $V_{DC} < E_{SY}/e = 2.6$ V ($e$ is the elementary charge). Here, charge carrier injection is insignificant, and the general device structure stays unaltered (EDLs form, but the AM remains undoped). For $V > E_{SY}/e$, which corresponds to a voltage where the LEC emits light, cf. Figure 2, it transitions into a structure resembling a parallel (RC)$_{EZ}$ element, with R and C being the emission zone (EZ) resistance and capacitance. The EDLs become conductive, and $C_{EDL}$ is now bypassed by a small charge-transfer resistance.[34] The EC-doped layers evolve, the thickness of the intrinsic OSC region shrinks, and the emission zone (EZ) forms between the p- and n-doped regions. As a result, the capacitance density of the LEC at high and intermediate frequencies changes from $C_{geo}$ to $C_{EZ}$, the capacitance of the undoped EZ, cf. Figure 4(c).[21,35] This transition of $C_{geo}$ into $C_{EZ}$ illustrates the LEC's doping process. The low-frequency impedance is dominated by the resistance of the undoped EZ, which now limits the DC in the device.

The D-OLED characteristics, cf. Figures 4(d-f), also resemble a transition into the characteristics of a parallel (RC) element with increasing $V_{DC}$. But there are three notable differences compared to the LEC: First, the low-frequency values for $|Z|$ and $\varphi$ start to decrease already for $V_{DC} < E_{SY}/e$. This implies that the injection-facilitating dipole reorientation of TMPE-OH is E-field driven rather than chemically induced following device fabrication. Second, for $V_{DC} > E_{SY}/e$, the sample does not show any change in $C_{geo}$. This proves that no doping and thus no p-i-n structure formation occurs. The AM remains undoped since it contains no ions. Third, for $V_{DC}$ = 3.8 V, the low-frequency (or DC) resistance is about 10 times higher than for the M-LEC. This can be rationalized by the thickness difference of the undoped (DC limiting) region in both devices: While the LEC DC is limited by the EZ resistance (the EZ width is on the order of 0.1 $d_{AM}$),[40,42] the DC in the D-OLED is limited by the entire $d_{AM}$.

To explain the significantly lower driving voltage of the D-OLED compared to the N-OLED observed in Figure 2, we can conclude from the IS experiments that it must be the injection resistance that is significantly reduced. IS showed that the D-OLED's AM exhibits a dipolar relaxation and that the device's DC impedance drops with increasing E-field while its high-frequency impedance remains unchanged, indicating that the dipole orientation facilitates bipolar charge injection into the semiconductor.

## OLED characterization

Now that we have established how dipolar doping achieves bipolar injection, we test the generality of this strategy. For that purpose, we use four different polar molecular species blended into SY to fabricate D-OLEDs using the same device stack as before. Figure S3 displays their voltage transients under constant-current operation for increasing doping concentration. We observe the same qualitative trend for all four dopants: With increasing dipole concentration, the driving voltage decreases, indicating dipole-mitigated injection for all four cases. When the doping level exceeds 50 - 60 wt%, however, the devices become prone to short-circuits, exhibit increased resistance, and are mostly non-functional. Such high dipolar doping levels yield poor film homogeneity and phase-separated domains, which potentially alter the SY packing and impair intermolecular charge-carrier transport.

All investigated dipolar dopants yield functional D-OLEDs with drastically enhanced forward luminance and reduced driving voltage, cf. Figures S4(b-d), relative to the N-OLED shown in Figure S4(a). Their optoelectronic performance depends on the properties of the dopants. While we initially chose TMPE-OH ($M_n$ = 1014 g mol$^{-1}$) whose HOMO/LUMO levels are outside the SY levels, cf. Figure 1(b), the lower-molecular-weight TMPE-OH ($M_n$ = 450 g mol$^{-1}$), for which we measure HOMO/LUMO energies of -5.3 eV and -2.9 eV, respectively, appears to introduce traps in SY that improve charge-carrier balance but also reduce mobility. This leads to higher forward luminance and slightly increased driving voltage. Please refer to Figures S4(b,c) for this comparison. As a further test, we replace SY with the sky-blue thermally activated delayed fluorescence (TADF) emitter 4TCzBN and dope it with TMPE-OH ($M_n \approx$ 450 g mol$^{-1}$), using the same device architecture as before. Again, we find decreasing driving voltages with increasing doping concentrations, cf. SI Section S6.

Finally, we compare the performance of the herein investigated D-OLED concept with SY-LECs and SY-OLEDs using dedicated injection layers. Figure 5 shows the performance of the N-OLED ($d_{AM}$ = 116 ± 5 nm) and the D-OLED ($d_{AM}$ = 132 ± 5 nm) comprising TMPE-OH ($M_n$ = 450 g mol$^{-1}$, 40 wt%) as the dopant. The current density-voltage-luminance (*JVL*) scans are conducted from 0 V to 7 V and back to 2 V in voltage steps of 0.1 V, each step taking 1 s, after pre-biasing the OLEDs for 10 s at 3 V. The N-OLED suffers from substantial injection resistance and shows no perceivable luminance beyond the sensitivity threshold of our setup, which is in agreement with the observations presented in Figure 2. It also shows no significant *J-V* hysteresis, consistent with the IS data in Figure 3(a-c), where we found no relaxation processes for the N-OLED. By blending TMPE-OH into the system, cf. Figure 5(b), we achieve a D-OLED with greatly enhanced current and luminance performance, yielding low-voltage operation and a current efficacy above 10 cd A$^{-1}$ at 100 cd m$^{-2}$ during the return scan, cf. Figure 5(c). The E-field dependence of the dipole orientation (the relaxation process found in Figure 3) introduces hysteresis in both current and luminance.

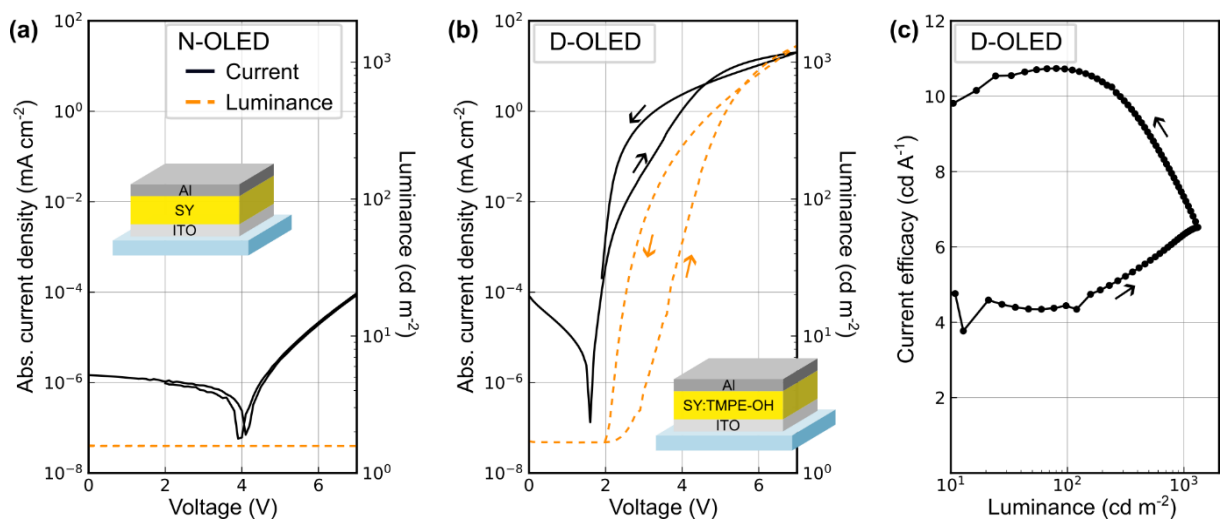

*Figure 5. JVL characterization of (a) the N-OLED and (b) D-OLED comprising a blend of SY:TMPE-OH ($M_n$ = 450 g mol$^{-1}$, 40 wt%) as AM. (c) Current efficacy of the D-OLED.*

This means that dipolar doping is a volatile, dynamic effect. It is therefore important to stress that the *JVL* characteristics of the D-OLED inherently depend on the measurement protocol and device history. Nevertheless, these characteristics allow a meaningful comparison with previously reported SY devices: Niu, Blom, and coworkers reported a SY OLED comprising PEDOT:PSS and Ba as injection layers and a 120 nm thick SY film, achieving a current density of about 5 mA cm$^{-2}$ at 5 V.[30] Burns, Yambem et al. presented a similar stack reaching about 1000 cd m$^{-2}$ at 10 mA cm$^{-2}$, and a current efficacy of 10 cd A$^{-1}$.[43] For SY LECs, Diethelm, Hany, and coworkers reported similar current efficacies.[28] In our measurement, we reach approximately 5 and 7 mA cm$^{-2}$ at 5 V and about 800 cd cm$^{-2}$ at 10 mA cm$^{-2}$. Thus, our *JVL* values are comparable to those of SY-OLEDs featuring injection layers and SY-LECs.

## Discussion

Concerning the dipolar compound, it may seem counterintuitive that large PEG-based polymers can reorient within a SY matrix, cf. SI Section S5. Instead of the entire polymer chain, however, only small segments like the ether (-C-O-C-) and hydroxyl (-O-H) groups may respond to the field. This reorientation still requires sufficiently large voids within the surrounding material and suggests that a soft, open, rubbery AM morphology is preferable to a dense, glassy, or crystalline one. Consequently, both the choice of the OSC and the AM's deposition and annealing conditions likely play an important role in enabling E-field-induced dipole injection.

We further observe a notable increase in driving voltage for the D-OLED over time, cf. Figures 2(a) and S4(b-d). As studied for PEO-based systems,[44,45] the terminal OH groups in TMPE-OH or the ether backbone are susceptible to oxidation at relatively low potentials.[46,47] TMPE-OH may thus act as a Brønsted acid, and deprotonation of the OH groups can produce hydrogen ions, leading to device degradation. We therefore suggest exploring other aprotic polar molecules.

Finally, the magnitude of the external E-field is an additional critical factor, since it exerts the torque that causes dipolar reorientation. In our devices, the initial magnitude of the E-field is on the order of 10-100 MV m$^{-1}$. This field will be partially screened when the dipoles in the AM orient and thereby form uncompensated charged layers at the two electrode interfaces, similar to EDLs in electrochemical devices or films that exhibit spontaneous orientation polarization.[26,48] However, a fundamental distinction of the herein reported strategy is that it does not depend on a long-range motion of bulky ions, avoids high voltages during fabrication,[23,24] and is instead activated and controlled by the E-field generated by the applied operating voltage.

# Conclusion

We report a method for achieving bipolar charge injection from air-stable electrodes into a single organic film. This is achieved by blending an auxiliary dipolar compound into the OSC, a process referred to as dipolar doping. Under operation bias, these dipoles reorient in response to the applied E-field and yield injection-assisting EDLs at the electrode interfaces. Dipolar doping is advantageous for solution-based fabrication under ambient conditions because it is compatible with air-stable electrode materials, does not rely on the dipole orientation of the OSC or on mobile ions that can compromise device performance, avoids high voltages during fabrication, and eliminates the need for additional injection layers. We demonstrate the merits of this concept by fabricating a single-layer dipole OLED comprising the electroluminescent polymer SY and the dipolar compound TMPE-OH, which operates comparably to established multi-layer SY-OLEDs and LECs. We further show that efficient dipolar doping can be achieved with multiple dipolar molecules and a different emitter material. Thus, our approach is both practical and broadly applicable.

## Experimental

### Ink fabrication

The AM constituents are a phenyl-substituted poly(para-phenylenevinylene) copolymer termed "Super Yellow" (SY, Livilux PDY-132, Merck, Germany), a hydroxyl end-capped trimethylolpropane ethoxylate (TMPE–OH, $M_n$ = 1014 g mol$^{-1}$, Merck, Germany), and a KCF$_3$SO$_3$ salt (Solvionic, France). The salt is dried in a vacuum oven at $p$ < 1 mbar and 190 °C for 12 h and filtered with a 0.1 μm PTFE filter (Cytiva, Puradisc 25, Whatman) before use. TMPE-OH is dried in a vacuum oven at $p$ < 1 mbar and 50 °C for 12 h. For the AM ink formulation, the constituents are separately dissolved in cyclohexanone (Sigma-Aldrich, USA) at a concentration of 10 mg ml$^{-1}$ (SY), 20 mg ml$^{-1}$ (TMPE–OH), and 10 mg ml$^{-1}$ (KCF$_3$SO$_3$). They are stirred at 70 °C on a magnetic hot plate positioned in a glovebox ([O$_2$] < 1 ppm, [H$_2$O] < 1 ppm) for ≥ 24 h. For the formulation of the five AM inks, these master inks are blended in five different solute mass ratios (cf. Table 1). To yield a SY concentration of 7.25 mg ml$^{-1}$ in the final ink, which determines the resulting AM film thickness, cyclohexanone is added accordingly. The final AM inks are then stirred again under the same conditions for ≥ 24 h. For the D-OLED presented in Figure 5, a different TMPE–OH ($M_n$ = 450 g mol$^{-1}$, Sigma-Aldrich, USA) is used.

### Device fabrication

The indium tin oxide (ITO) coated (thickness = 145 nm, $R_s$ = 20 Ω sq$^{-1}$) glass substrates (substrate area = 30 × 30 mm$^2$, thickness = 0.7 mm, Kintec, HK) are cleaned in an ultrasonic bath by sequentially using a detergent (Extran MA 01, Merck, GER) in deionized water, deionized water, acetone (VWR, GER), and isopropanol (VWR, GER). The ITO-coated substrates are dried in an oven at 120 °C for ≥ 12 h. The AM ink is spin-coated on the substrate (2000 rpm, acceleration = 2000 rpm s$^{-1}$, time = 60 s) and dried on a hot plate at 70 °C for 1 h. The AM thickness ($d_{AM}$) is measured with a stylus profilometer (Dektak XT, Bruker, USA). The values for $d_{AM}$ are summarized in Table 1. The reflective Al top electrode (thickness = 100 nm) is deposited by thermal evaporation at a base pressure $p < 6 \cdot 10^{-6}$ mbar, with a shadow mask defining the cathode area. The spatial overlap between the cathode and the anode defines four 2 × 2 mm² LEC pixels on each substrate. For the IS and OLED $JVL$ measurements, the devices are encapsulated with a cover glass (24 × 24 mm$^2$, VWR, GER) using a UV-curable epoxy resin (Ossila, UK) and measured under ambient conditions.

### Device characterization

The voltage-luminance transients are recorded inside an N$_2$-filled glovebox ([O$_2$], [H$_2$O] < 1 ppm, T ≈ 22 °C). The devices are driven by a constant current of 0.31 mA, corresponding to a current density of $J$ = 7.7 mA cm$^{-2}$, with the voltage compliance set to 21 V. The luminance is measured with a calibrated photodiode (S9219-01, Hamamatsu Photonics), and the voltage is tracked using a source measure unit (U2722A, Agilent).

The IS are recorded with the devices placed on a custom-built temperature stage to ensure a stabilized temperature of $T$ = 25 ± 0.02 °C. The temperature stage comprises a Peltier element, powered by a DC power supply (E3631A Agilent) and monitored by a PT100 temperature sensor read by a multimeter (34401A Agilent). The feedback between the Peltier power supply and the temperature sensor is controlled using a Python-based PID controller.

The IS data are recorded by a potentiostat equipped with a frequency analyzer (Metrohm Autolab PGSTAT302). Before starting a frequency scan (100 kHz → 10 mHz → 100 kHz, 10 steps per decade, $V_{AC}$ RMS = 20 mV), the devices are stabilized at $T$ = 25 ± 0.02 °C for 120 s to ensure thermal equilibrium. If $V_{DC}$ > 0 is used, the devices remain biased for 1200 s before starting the measurement to ensure steady-state conditions. A few data points (mostly around the net frequency $f$ = 50 Hz) exhibit unreasonable phase jumps and have been removed for clarity. Both the frequency analyzer and the temperature stage are controlled by SweepMe! ([www.sweep-me.net](www.sweep-me.net)), a multitool measurement software ensuring fully automated measurements.

For the OLED *JVL* characteristics, current and voltage are measured using a source measure unit (SMU, Keithley 2400) in a two-wire configuration. For the luminance data, a second SMU (Keithley 2400) measures the photocurrent generated by a photodiode (BPW21, Osram), equipped with an eye-response filter and calibrated using a luminance meter (Konica Minolta LS-110). All instruments are controlled by SweepMe!

## Equivalent circuit model

The ECM displayed in Figure 3(d) is modeled in Python using LTspice (Linear Technology), which contains a self-consistent, numerical frequency-domain solver for AC analysis. The AC amplitude is set to RMS = 20 mV, i.e. the same as in the measurement. The fixed parameter values for $C_{geo}$ and $C_{EDL}$ are directly taken from the experiment (Table 2). The free parameter $R_{bulk}$ is chosen to match the experimental data. The full parameter set is given in the SI Section S4, and the Python script is available for download.

## Data availability

All relevant data, SweepMe! setting files, and the Python script running the ECM AC analysis are available for download here: paste link after acceptance

## Acknowledgements

The authors acknowledge generous financial support from the Swedish Research Council (2019-02345 and 2021-04778), Kempe Foundations, Bertil och Britt Svenssons stiftelse för belysningsteknik, the Knut and Alice Wallenberg Foundation for a Proof-of-concept grant (KAW 2024.0497), and the Wallenberg Initiative Materials Science for Sustainability (WISE) funded by the Knut and Alice Wallenberg Foundation (WISE-AP01-D02 and WISE-AP02-PD21). L.E. acknowledges financial support from the European Union through an ERC Advanced Grant (ERC, InnovaLEC, 101096650). A.K. acknowledges funding from the European Union (HORIZON MSCA 2023 PF, acronym UNID, grant number 101150699). A.K. thanks Louis Conrad Winkler and Dr. Jakob Wolansky (both TU Dresden, Germany) for their support with upgrading the Autolab SweepMe! driver. J.R.R. and A.K. thank Mikael Fredriksson and Peter Wikström for their support with crafting the temperature stage in the workshop. N.M. and A.K.P. acknowledge financial support from the Swedish Research Council (Grant No. 2021-05784), the Knut and Alice Wallenberg Foundation through the Wallenberg Academy Fellows Programme (Grant No. 2023.0089), and Kempestiftelserna (WISE-UmU/Kempe-program, Grant No. JCSMK 23-198).

# Supporting Information

## In-operando dipole orientation for bipolar injection from air-stable electrodes into organic semiconductors

Anton Kirch[1*], Joan Ràfols-Ribé[1], Kumar Saumya[1], Thushar Salkod Mahabaleshwar[1,3], William Strömberg[1], Ajay Kumar Poonia[1,2,3], Preetam Dacha[1], Yuntao Qiu[1], Sri Harish Kumar Paleti[1,3], Christian Larsen[1], Nicolò Maccaferri[2,3], and Ludvig Edman[1,3]

[1] The Organic Photonics and Electronics Group, Department of Physics, Umeå University, SE-90187 Umeå, Sweden

[2] Ultrafast Nanoscience Group, Department of Physics, Umeå University, SE-90187 Umeå, Sweden

[3] Wallenberg Initiative Materials Science for Sustainability, Department of Physics, Umeå University, SE-90187 Umeå, Sweden

* Correspondence: anton.kirch@umu.se

## Table of contents

# S1. Cyclic voltammetry measurements

We perform cyclic voltammetry (CV) measurements to extract the energy levels of Super Yellow and TMPE-OH on a Metrohm PGSTAT302 using a standard electrochemical cell at a scan rate of 50 mV/s. Super Yellow's CV is measured with a thin film of SY on gold-coated glass electrodes as the working electrode (WE). The thin films are prepared through spin coating an 8 g/L solution in cyclohexanone at 2000 rpm (2000 rpm/s acceleration) for 60 s and annealed for 1 h at 70 °C. TMPE-OH's CV is measured in 0.2 M ACN solution with a gold-coated glass as the WE. Both measurements use a Pt wire as counter electrode (CE), Ag wire as quasi reference electrode (RE), and 0.1 M Tetrabutylammonium hexafluorophosphate (TBAPF$_6$, > 99.0 %, Merck)/ACN (anhydrous, Fischer Scientific) as the supporting electrolyte. The measurements are calibrated to the Ferrocene (Fc/Fc$^+$) redox couple and are performed in an inert environment (< 0.5 ppm O$_2$, < 0.5 ppm H$_2$O).

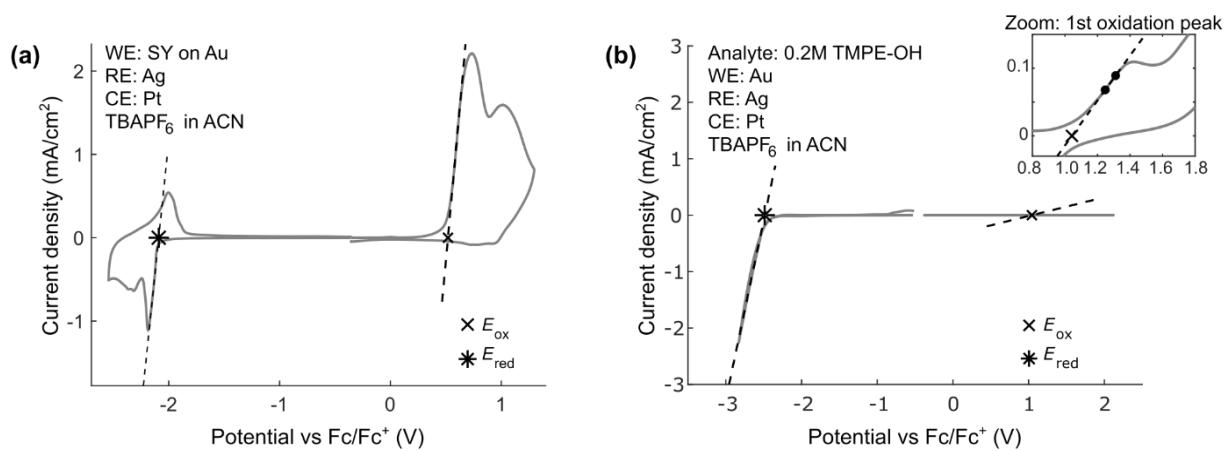

Figure S1. Cyclic voltammograms of (a) SY and (b) TMPE-OH ($M_n$ = 1014 g mol$^{-1}$) to extract their redox potentials.

The HOMO and LUMO levels we find are -5.3 and -2.7 eV for SY, -5.8 and -2.3 eV for TMPE-OH ($M_n$ = 1k g/mol), and -5.3 and -2.9 eV for TMPE-OH ($M_n$ = 450 g/mol). They are calculated from the voltage onset potentials derived from tangents at the half-maximum current using the relation:

$$E_{\text{HOMO/LUMO}} = -e(4.8 + E_{1/2}),$$

where $E$ is the HOMO/LUMO level in eV, $e$ is the fundamental electronic charge, and $E_{1/2}$ is the voltage onset potential vs. Fc/Fc+ in V. (J. Pommerehne, et al. "Efficient two layer leds on a polymer blend basis." *Advanced Materials* 7.6 (1995): 551-554.)

## S2. Voltage-luminance turn-on transients

Figure S2 displays the turn-on characteristics during the first 10 s of operation for the three LECs (open symbols) and the D-OLED (closed squares). The pristine devices are identically fabricated as those in the main manuscript and driven at a constant current of 1 mA (corresponding to 25 mA/cm$^2$).

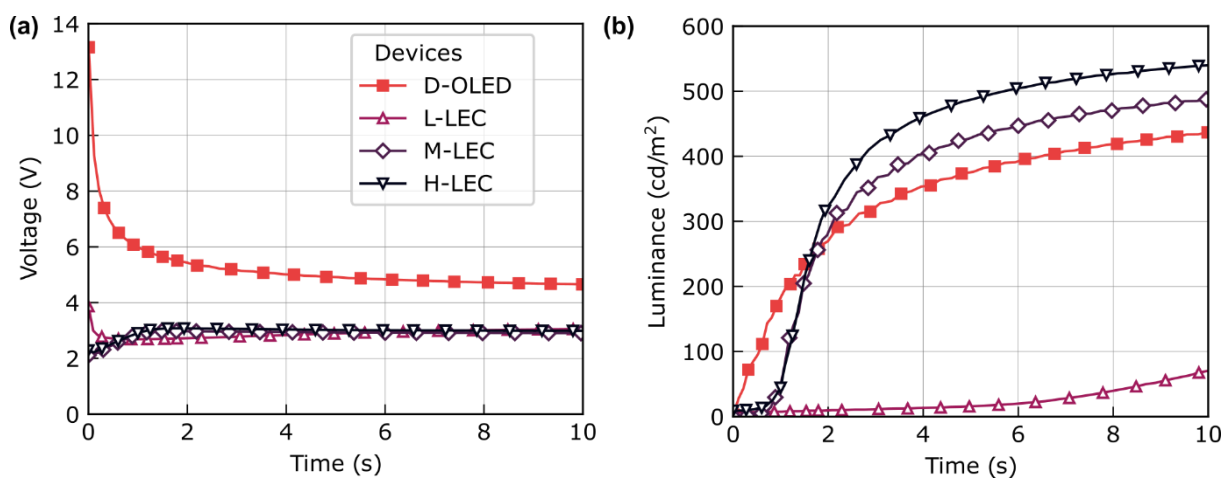

*Figure S2. Temporal evolution of (a) the driving voltage and (b) the forward luminance for devices driven by a constant current of 1 mA (= 25 mA/cm$^2$) using a voltage compliance of 21 V.*

Here, we use an Agilent U2722A SMU for driving the devices. The luminance is measured with a calibrated photodiode (S9219-01, Hamamatsu Photonics) and is read by the same Agilent SMU. In contrast to the data presented in the main manuscript, Figure 2, the data acquisition is handled by a custom-made Python program, which allows us to record the voltage-luminance transients at 100 ms intervals.

## S3. Capacitance density

IS measures the quantities $|Z|$ and $\varphi$, corresponding to the real and imaginary parts of Z ($R$ and $X$, respectively), with $i$ being the imaginary unit:

$$Z = R + iX,$$
$$R = \text{Re}(Z) = |Z|\cos(\varphi),$$
$$X = \text{Im}(Z) = |Z|\sin(\varphi).$$

The capacitance $C'$ is defined via the admittance, the inverse of the impedance. The capacitance density $C = C'/A$ considers the pixel area $A$.

$$C' = \frac{1}{\omega}\text{Im}\left(\frac{1}{Z}\right)$$

$$= \frac{1}{\omega}\text{Im}\left(\frac{1}{R+iX}\right) = \frac{1}{\omega}\text{Im}\left(\frac{(R-iX)}{(R+iX)(R-iX)}\right)$$

$$= \frac{1}{\omega}\text{Im}\left(\frac{R-iX}{R^2+X^2}\right) = \frac{-X}{\omega|Z|^2}$$

$$C = \frac{C'}{A} = \frac{-X}{A\,\omega|Z|^2}$$

# S4. Equivalent circuit model (ECM) parameters

The ECM analyzes the Debye circuit displayed in the inset of Figure 3(d) in the main manuscript using LTspice. The netlist of the ECM is evaluated in a Python script using the LTspice package and running LTspice in the background. It looks as follows:

```
V1 0 N002 AC 0.02
C1 0 N002 {C1}
R1 0 N001 {R1}
C2 N001 N002 {C2}
.ac dec 10 1e-2 1e5
.end
```

V1 corresponds to the applied $V_{AC}$, C1 to $C_{geo}$, C2 to $C_{EDL}$, and R1 to $R_{bulk}$. C1 and C2 are used according to the measured values (cf. Table 2 in the main manuscript). R1 is the only fitting parameter and was chosen such that the modeled IS data resembled the experimental data.

*Table S3. Parameter values used for the ECM in the main manuscript.*

|  | N-OLED | D-OLED | L-LEC | M-LEC | H-LEC |
|---|---|---|---|---|---|
| {C1} (F) | 9.14e-10 | 9.96e-10 | 1.19e-9 | 1.51e-9 | 1.60e-9 |
| {R1} (Ω) | 2e11 | 8e9 | 3e7 | 8e6 | 3e6 |
| {C2} (F) | 1.3e4 | 1.3e4 | 9.11e-8 | 1.27e-7 | 1.27e-7 |

The EDL capacitances (C2) of the N-OLED and D-OLED should be infinite, since they do not comprise mobile ions. As a numerical representation, we put them at extremely high values.

The full Python code is available for download. Please find the link in the main manuscript under "Data availability".

## S5. Further polar molecules as dopants

We blend four different polar organic molecules into SY, fabricate devices as in the main manuscript (ITO/SY:polar molecule/Al) with $d_{AM}$ = 40 (± 5) nm, and run them at a constant current of 1 mA (= 25 mA/cm$^2$). The polar molecules are:

a) TMPE-OH, $M_n$ ≈ 450 g/mol (Sigma-Aldrich, USA)
b) TMPE-OH, $M_n$ ≈ 1014 g/mol (Merck, Germany)
c) PEGDME: Poly(ethylene glycol) dimethyl ether, $M_n$ ≈ 1000 g/mol (Sigma-Aldrich, USA)
d) PEG-PPG-PEG: Poly(ethylene glycol)-block-poly(propylene glycol)-block-poly(ethylene glycol), $M_n$ ≈ 14600 g/mol (Sigma-Aldrich, USA)

For all devices, we observe that increasing the concentration of polar molecules reduces the driving voltage. Beyond 40 wt%, the AM becomes non-uniform, the driving voltage increases again, and the devices become prone to shorts. The data of the non-functional devices is not shown for clarity.

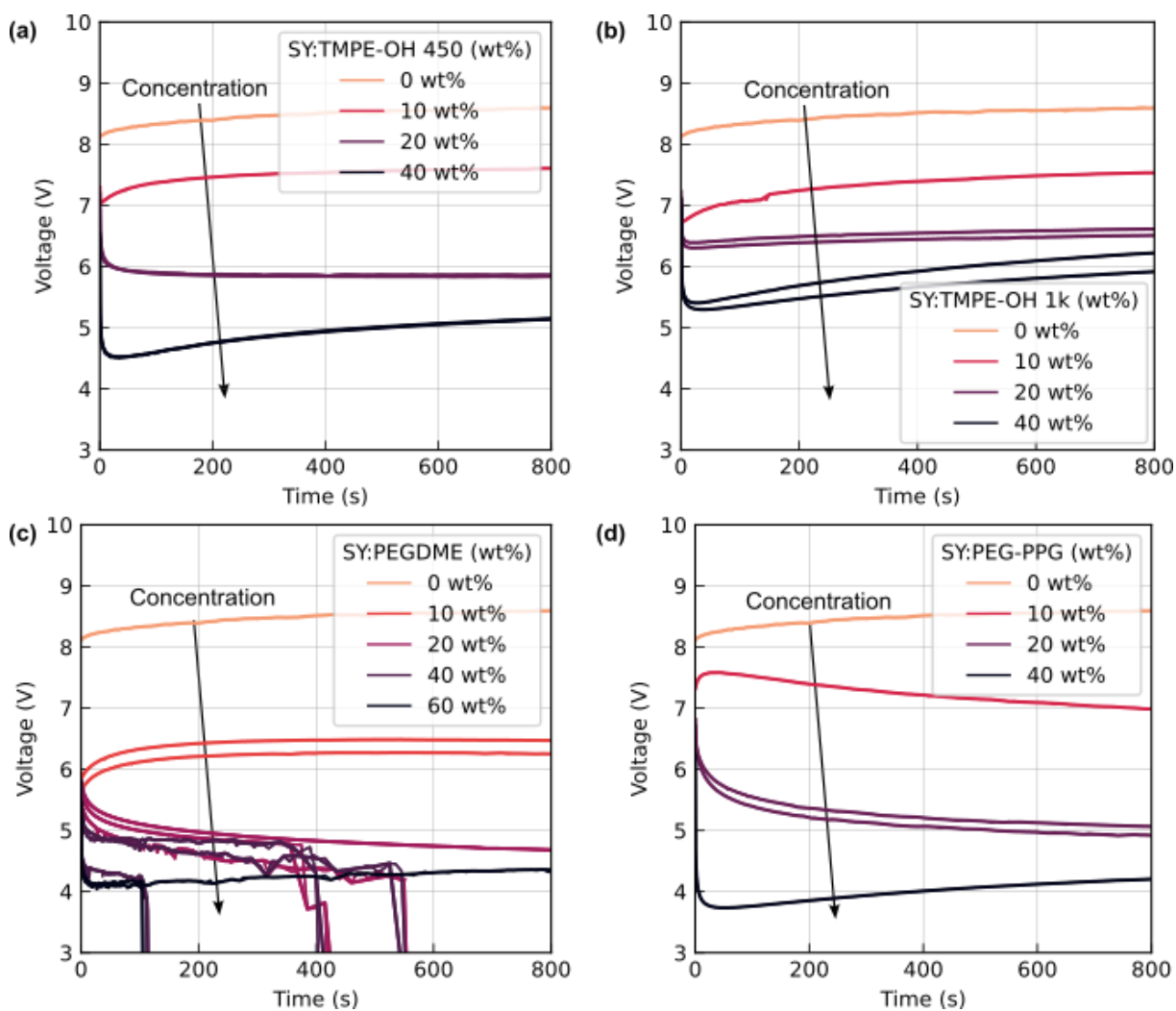

*Figure S3. Dependence of the driving voltage on the concentration of (a) TMPE-OH, $M_n$ ≈ 450 g mol$^{-1}$, (b) TMPE-OH, $M_n$ ≈ 1014 g/mol, (c) PEGDME, and (d) PEG-PPG-PEG. The pristine devices are driven at 1 mA (= 25mA/cm$^2$).*

Now, we fabricate D-OLEDs using the above dipolar compounds and test their voltage-luminance transients compared to the N-OLED characteristics. Their film thickness is increased compared to the previous measurement to achieve bright devices, cf. Table S2.

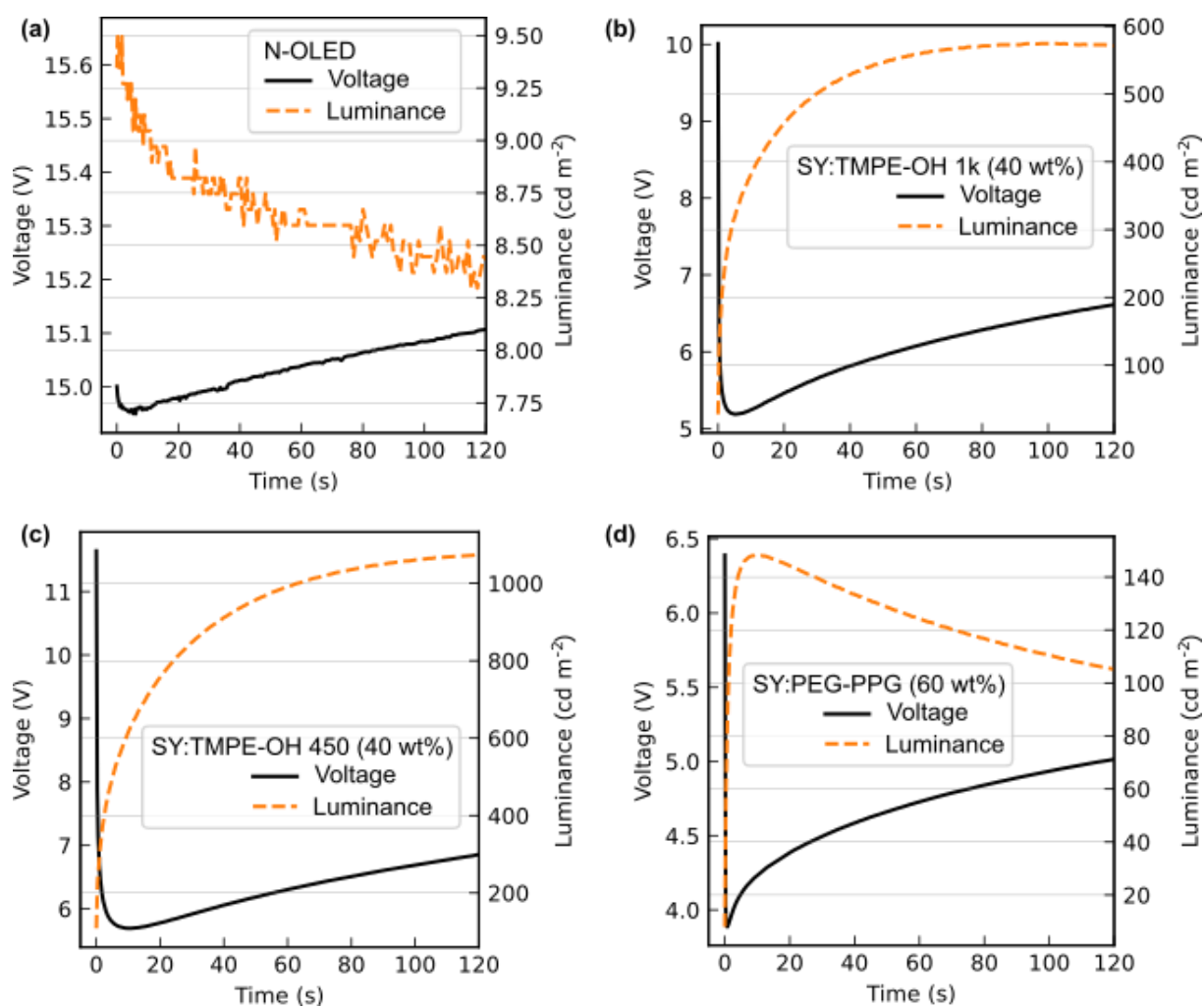

*Figure S4. Voltage-luminance transients for devices driven at 0.4 mA (10 mA/cm$^2$). Subfigure (a) displays the N-OLED characteristics, and (b-d) contain the dipole compound indicated in the legends.*

Figure S4 shows that all auxiliary dipolar compounds significantly decrease the driving voltage and increase the forward luminance, compared to the N-OLED presented in Figure S4(a).

Table S4. Active layer thickness ($d_{AM}$) of the devices presented in Figure S4.

| Device | N-OLED | D-OLED | | |
|---|---|---|---|---|
| | | TMPE-OH 450 | TMPE-OH 1k | PEG-PPG |
| $d_{AM}$ (nm) | 117 ± 5 | 132 ± 5 | 134 ± 5 | 148 ± 10 |

## S6. Using the TADF emitter 4TCzBN instead of SY

In this experiment, we replace SY with the sky-blue thermally activated delayed fluorescence (TADF) emitter 4TCzBN and dope it with TMPE-OH ($M_n \approx 450$ g/mol) at increasing weight percentages. As observed for the SY devices (Figure S3), an increasing dipolar doping reduces the driving voltage significantly. The luminance, on the other hand, decreases with increasing TMPE-OH concentration. This reflects that the effectiveness of dipolar doping depends on the interplay of host and dopant, altering the electron and hole trap density, conductivity, and film morphology. For a small-molecule emitter, a polymer dopant like PEG-PPG-PEG may be better suited to produce homogeneous films.

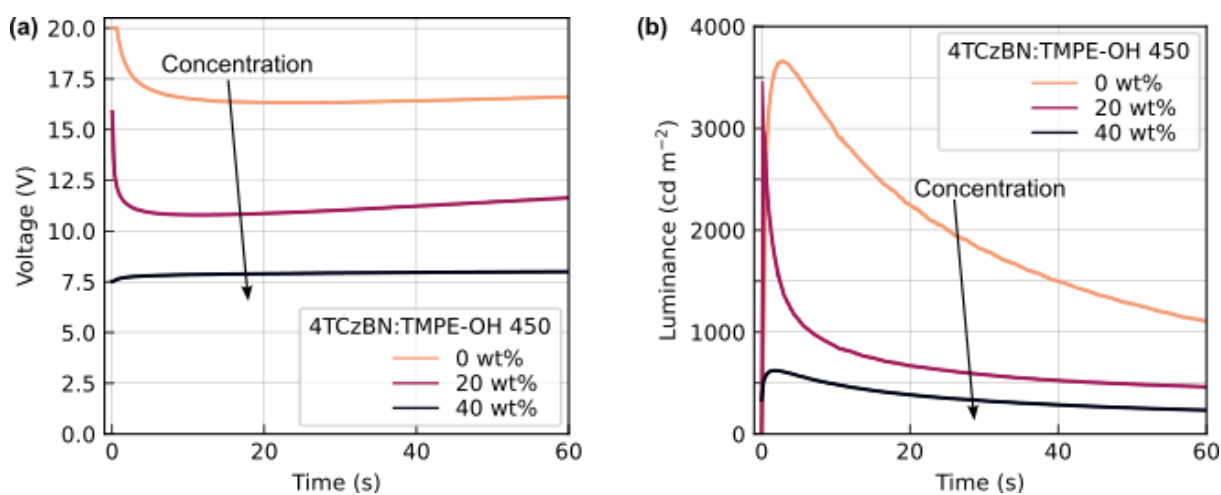

*Figure S5. (a) Voltage and (b) luminance transients for ITO/AM/Al devices driven at a constant current of 1 mA (25 mA cm$^{-1}$). The AM consists of TMPE-OH (Mn ≈ 450 g/mol) blended into 4TCzBN at the indicated weight percentage.*

The fabrication of these devices is equivalent to the fabrication of the devices used in the main manuscript, only that SY is replaced by the TADF emitter 2,3,5,6-Tetrakis[3,6-bis(1,1-dimethylethyl)-9H-carbazol-9-yl]benzonitrile (4TCzBN, Lumtec, Taiwan). This material is dried in a vacuum oven at p < 1 mbar and 65 °C for 24 h, dissolved in Chlorobenzene at a concentration of 40 mg/ml, and stirred at 70 °C overnight on a hotplate. The TMPE-OH ink is prepared as in the main manuscript and added according to the desired weight percentage. This AM ink is filtered with a 0.45 μm PTFE filter (VWR) and spin-coated under the same conditions as the SY devices. The $d_{AM}$ of the dipole TADF OLEDs is displayed in Table S3. We probe the film thickness at several points close to the pixels and see that the film homogeneity (reflected by the uncertainty) decreases with increasing dipolar doping.

*Table S5. AM thickness for the investigated TADF D-OLEDs consisting of 4TCzBN:TMPE-OH (x wt%).*

| Device      | 0 wt%   | 20 wt%   | 40 wt%   |
|-------------|---------|----------|----------|
| $d_{AM}$ (nm) | 163 ± 5 | 158 ± 10 | 145 ± 15 |